\documentstyle[12pt]{article}
\begin{document}
\setlength{\baselineskip}{20pt}
\begin{center}
{\Large{\bf Spherical symmetric charged solution with cosmological constant}}
\end{center}
\begin{center}
N. \"Ozdemir\\
ITU Faculty of Science and Letter\\
34469, Istanbul, Turkey\\
and\\
Feza G\"ursey Institute\\
Emek Mah. Rasatane Yolu  No: 68,\\ 34684 \c Cengelk\"oy, Istanbul, Turkey
\end{center}
\vfill
\begin{abstract}
\noindent A spherically symmetric charged ideal fluid solution of Einstein field equation
is given in the presence of the cosmological constant and two well known example of this
type of solution is presented. If the matter is confined in a region, the exterior
spacetime is considered as RN-de Sitter (Reissner-Nordstr\" om de Sitter) and to complete
solution matching conditions are examined. We show that the function which is related to
the dynamics  of the system will determine the fate of the system: expansion, contraction
or bouncing situations may occur for different configurations. The initial conditions of
the matter determine the final form of the system and therefore the nature of the
singularities in the presence of the electric charge and the cosmological constant is
examined to reveal their effects on the singularity formation during collapse.

\vspace{5mm} \noindent PACS number(s):  04.70.Bw, 04.20.Dw, 04.20.-q

\end{abstract}

\section{Introduction}
Cosmological observations indicate that the expansion of the universe is accelerating. A
simple phenomenological interpretation of the data in terms of dark energy has been
successful thus far \cite{peebles}; dark energy is assumed to have negative pressure,
unlike ordinary mass-energy, and thus leads to a negative force that may account for the
acceleration of the universe. Recent observations indicate that universe is dominated by
dark energy ($\sim 70 \% $), which can be thought of as a perfect fluid with an
energy-momentum tensor given by
\begin{equation}
T^{\mu\nu}_D=(\mu_D+p_D)\,u^\mu u^\nu+p_D\, g^{\mu\nu}, \label{fluid}
\end{equation}
where $\mu_D+p_D=0$ and $p_D< 0$. The inclusion of source term (\ref{fluid}) in
Einstein's field equations amounts simply to a cosmological constant $\Lambda$ given by
\\$\Lambda=- p_D>0$. In this way, dark energy may be represented by a positive
cosmological constant $\Lambda$ in the Einstein field equations
\begin{equation}
R_{\mu\nu}-{1\over 2}g_{\mu\nu} R+\Lambda g_{\mu\nu}= T_{\mu\nu}. \label{gen1}
\end{equation}

The cosmological constant has a long history in Newtonian and relativistic gravity
theories. It was introduced into general relativity by Einstein as a means of balancing
the gravitational attraction of the matter on cosmological scales, leading to the
Einstein static universe model. Alternatively, it can be thought of as a measure of the
energy density of the vacuum.

In this paper the gravitational motion of charged matter is considered in the presence of
a cosmological constant. To simplify matters, \textit{a perfect fluid} distribution is
considered that is electrically charged and undergoes spherically symmetric collapse or
expansion in the absence of external fields. The exterior field is given by the
Reissner-Nordstr\" om-de Sitter metric
\begin{equation}
ds^2=-(1-{2 M\over \hat{r}}+{Q^2\over \hat{r}^2}-{\Lambda \over
3}\,\hat{r}^2)d\hat{t}^2+{d\hat{r}^2\over (\displaystyle{1-{2 M\over \hat{r}}+{Q^2\over
\hat{r}^2}-{\Lambda \over 3}\,
\hat{r}^2})}+\,\hat{r}^2(d\hat{\theta}^2+\sin^2\hat{\theta} d\phi^2)\label{mmm1}
\end{equation}
where $M$ is the net mass and $Q$ is the net charge of the system and $\Lambda$ is the
cosmological constant of the RN-de Sitter spacetime.

%In the cosmological point of view, the ``scalar'' cosmological constant can either be
%constant or time-varying or coordinate-dependent. The time-dependent cosmological
%constant idea is consistent with the Einstein's field equations in which the matter part
%is created or destroyed to compensate for it \cite{aldrovandi}. A coordinate-dependent
%cosmological constant example is given in \cite{ray} where the spacetime needs to have an
%anisotropic charged source to get spherically symmetric solution of Einstein-Maxwell
%field equations.

The gravitational collapse phenomena is still open problem in the general relativity and
it has not been taken its final form yet. Throughout the studies it is shown that the
initial conditions are determinative of the final fate of the collapse \cite{joshi}. If
the collapse can not be stopped in an equilibrium state and allows to formation of the
singularities as a result, the end state of the collapse can either be black hole or
naked singularity depending on the character of the singularities. If all singularities
are hidden behind an event horizon singularities can not be seen by distant observer and
the end state becomes "a black hole",  or if the singularities are bare and can be
visible by distant observer, the end state becomes "naked singularity" \cite{joshi}.

%By analyzing spherical symmetric gravitational collapse of any type of matter subject to
%the weak energy condition, it is shown that the formation of the naked singularity is
%related to choice of the initial data.
One of the physical features affected on the gravitational collapse is shear. In
\cite{joshi2}, shear effects on the gravitational collapse of the spherical massive cloud
with non-radial pressure are studied and shown that sufficiently strong shear effects
near singularity delay the formation of the apparent horizon and allow the formation of
the naked singularity.

In \cite{shapiro} effects of the cosmological constant on the gravitational collapse of
the pressureless matter is studied and shown that positive gravitational constant plays
repulsive role and slows down the collapse process.

As a property of matter how does electric charge affect the collapse phenomenon? This
question finds some answers through the following papers and needs to be completed. A
solution for the spherical symmetric charged stars are considered in point of view
formation of the black holes and voids in \cite{senovilla}. In \cite{lurdes}\,, the
gravitational collapse of the spherical symmetric charged radiating Vaidya-RN type
spacetimes are studied. To avoid singularities due to charge of the matter in spherical
symmetric collapse is examined in \cite{kras}. As a stellar model, relativistic
structure, stability and gravitational collapse of the charged fluid is studied in
\cite{ghezzi}, the specific values of the electric charge of the fluid allows formation
of naked singularity besides black hole formation. A spherically symmetric charged ideal
fluid is examined in \cite{mashhoon} due course of gravitational collapse. Similar to
solution presented in \cite{mashhoon}, in this work,  we give a solution of Einstein
field equations in de Sitter spacetime which is in isotropic form. Furthermore, we
examine the effects of the electric charge on the collapse phenomenon in the presence of
the cosmological constant.

In the following section, a spherically symmetric solution of the Einstein-Maxwell
equations is given in the presence of the cosmological constant. The exterior spacetime
of the charged fluid sphere is considered the RN-de Sitter spacetime and matching
conditions about two distinct (interior and exterior) regions are examined in section 3.
Two well known examples of this type of solutions, RN-de Sitter and Mc Vittie-de Sitter
are given. Section 4 is devoted to the gravitational collapse of the charged fluid and
formation of the singularities. Since the co-moving character of the spacetime may give
coordinate dependent results, it is necessary to use coordinate free "null geodesics
method" where the nature of the singularities does not change its character. Therefore,
the nature of the singularities are investigated null geodesic method. Throughout the
paper we use units such that $c=1$ and $(8\pi\,G=1)$.

\section{Interior solution}
Imagine a co-moving system of coordinate for the interior $(t,
\rho,\theta, \phi)$ that remains at rest with the moving charged
matter and is given by
\begin{equation}
ds^2=-a^2\, dt^2+b^2\,d\rho^2+R^2\,\rho^2 d\theta^2+R^2\,\rho^2\sin^2\theta d\phi^2
\label{flrwline}
\end{equation}
where  $a=a(t,\rho)$, $b=b(t,\rho)$ and $R=R(t,\rho)$ are arbitrary positive functions of
time coordinate $t$ and radial coordinate $\rho$. To consider shear free motion of the
matter that is the ideal fluid, we take $R=\rho\,b(t,\rho)$. (\ref{flrwline}) is assumed
to be the solution of the field equations (\ref{gen1}) with a cosmological constant
$\Lambda_0$ and a source $T_{\mu\nu}=T^{\rm m}_{\mu\nu}+T^{\rm em}_{\mu\nu}$, where
\begin{equation}
T^{\rm m}_{\mu\nu}=(\mu+p)\,u_\mu u_\nu+p\, g_{\mu\nu}
\end{equation}
and the electromagnetic energy-momentum tensor is defined by
\begin{equation}
T^{\rm em}_{\mu\nu}=2(g^{\alpha\beta}F_{\alpha\mu}F_{\beta\nu}-{1\over 4}g_{\mu\nu}
F_{\alpha\beta}F^{\alpha\beta})\,. \label{enem}
\end{equation}
Here electromagnetic field tensor is $F_{\mu\nu}=\partial_\nu A_\mu-\partial_\mu A_\nu$
and $A_\mu$ is the vector potential. The spherical symmetry of the spacetime ensures
existence of a radial electric field in general.\\ After a suitable choice of the gauge
is $A_i=0$, $A_t=\Phi(t,\rho)$ chosen, according to the metric (\ref{flrwline}) the only
non-zero component of $F_{\mu\nu}$ becomes $F_{t\rho}=-\partial \Phi/\partial \rho$\,.
Then, non-vanishing components of the electromagnetic energy momentum tensor for the
spacetime with the gauge chosen above are given by
\begin{eqnarray}
T_t^t&=&-{q^2\over \rho^4\,b^4}\,,\nonumber\\[.2cm]
T_t^t&=&T_\rho^\rho=-T_\theta^\theta=-T_\phi^\phi
\end{eqnarray}
where $q$ is the total charge of the fluid. It is easily seen that $T=tr\,T_\mu^\nu=0$ as
expected. The Maxwell equations
\begin{equation}
{\partial\over\partial x^\mu}[\sqrt{-g}F^{\nu\mu}]=4\pi\sqrt{-g} J^\nu
\end{equation}
with $\sqrt{-g}=a b^3 \rho^2\sin\theta$ for the spacetime (\ref{flrwline}) become
\begin{equation}
{\partial\over\partial t}\left({b\over a}{\partial\Phi\over\partial \rho}\right)=0\,.
\end{equation}

It is seen that total charge $q$ defined by
\begin{equation} {b\over a}{\partial \Phi\over\partial\rho}={q\over
\rho^2}
\end{equation}
is independent of time and related to the charge density $\zeta$ by the equation
\begin{equation}
{dq \over d\rho}=4\pi \rho^2 b^3\zeta.
\end{equation}
The electric current is given  by $J^t=\zeta u^t$ in terms of the electric charge density
and 4-velocity $u^t=a^{-1}$. Since the total energy momentum tensor is the sum of the
electromagnetic $T^{\rm em}_{\mu\nu}$ and matter part $T^{\rm m}_{\mu\nu},$ the total
energy-momentum components will be obtained as
\begin{eqnarray}
&&{T_t}^t=-\mu-{q^2\over \rho^4\,b^4}\,,\qquad {T_\rho}^\rho=p-{q^2\over \rho^4\,b^4}\,,\nonumber\\[.2cm]
&&{T_\theta}^\theta=p+{q^2\over \rho^4\,b^4}\,,\qquad\,\,\, {T_\phi}^\phi=p+{q^2\over
\rho^4\,b^4}\, \label{enmom}
\end{eqnarray}
where $\mu$ are $p$  are  the matter-energy density and  pressure respectively. It is
supposed that the matter field satisfies the weak energy condition. For any timelike
vector $v^\mu$
\begin{equation}
T^{\rm m}_{\mu \nu}\,v^\mu\,v^\nu\,\geq\,0\end{equation} which gives
\begin{equation}
\mu\,\geq\,0,\qquad \mu+p\,\geq\,0\,.\label{enmomcon}
\end{equation}
Furthermore, the conservation of the energy $u_{\mu}T^{\mu\nu}_{\quad;\,\nu}=0$ gives the
relation
\begin{equation}
{1\over (\mu+p)}\,{\partial\mu\over \partial t}=-\,{3\over b}\,{\,\partial b\over
\partial t}\,\,. \label{encon}
\end{equation}

Now we are going to examine the geometry part of the problem given by the metric
(\ref{flrwline}) and consider the Einstein tensor $G_{\mu\nu}=R_{\mu\nu}-R\,
g_{\mu\nu}/2$ as follows
\begin{eqnarray}
G_{tt}&=&3\,{{\dot b}^2\over b^2}-{a^2\over b^2}\left[2\,{b^{\prime\prime}\over
b}-\left({b^\prime\over b}\right)^2+{4\over \rho}
\,{b^{\,\prime}\over b}\right]\nonumber\\[.2cm]
G_{t\rho}&=&-2 a\left({\dot{b}\over a b}\right)^\prime\nonumber\\[.2cm]
G_{\rho\rho}&=&\left({b^{\,\prime }\over b}\right)^2+2\,{a^\prime\over a}{b^\prime\over
b}+{2\over \rho}\,\left({b^\prime\over b}+{a^\prime\over a}\right)
-{b^2\over a^2}\left(2\,{\ddot {b}\over b}+{{\dot b}^2\over b^2}-2\,{{\dot a}\over a}\,{{\dot b}\over b}\right)\nonumber\\[.2cm]
G_{\theta\theta}&=&\rho^2\left[{1\over \rho}\left({a^\prime\over a}+{b^\prime\over
b}\right)+{a^{\prime\prime}\over a} +{b^{\prime\prime}\over b}-{b^{\prime 2}\over b^2}
+{b^2\over a^2}\left(2\,{\dot{a}\over a}{\dot{b}\over b}-{{\dot{b}}^2\over b^2}-2\,{\ddot{b}\over b}\right)\right]\nonumber\\[.2cm]
G_{\phi\phi}&=&\sin^2\theta\, G_{\theta\theta}.\label{fieldeq}
\end{eqnarray}
where ``dot'' and ``prime'' represent derivatives with respect to $t$ and $\rho$,
respectively.

By using the energy-momentum tensor of the fluid  obtained in (\ref{enmom}), the field
equations (\ref{gen1}) take the form
\begin{eqnarray}
G_{t\rho}&=&0\,,\label{bir}\\[.2cm]
{1\over a^2}\,G_{tt}\,-\Lambda_0&=&\mu+{q^2\over \rho^4\,b^4}\,, \label{alti}\\[.2cm]
{1\over b^2}\,G_{\rho\rho}\,+\Lambda_0&=&  p-{q^2\over \rho^4\,b^4}\,, \label{bes}\\[.2cm]
G_{\rho\rho}-{1\over \rho^2}\,G_{\theta\theta}&=&-\,{2\,q^2\over \rho^4\,b^2}\label{iki}
\end{eqnarray}
where $\Lambda_0$ is the cosmological constant for the interior region. For the sake of
the generality, the cosmological constant of the interior region $\Lambda_0\,$ is taken
to be different from the cosmological constant of the exterior region $\Lambda\, $ in the
beginning.

Then, the trace of field equations becomes
\begin{eqnarray}
(3p-\mu)-{2\over b^2}\left({a^{\prime\prime}\over a}+{a^\prime\, b^\prime \over a\,
b}-{b^{\prime  2} \over b^2}+{2b^{\prime\prime} \over b}+{2a^{\prime} \over a\,
\rho}+{4b^{\prime} \over b\, \rho}\right) -{6\over a^2}\left({\dot{a}\,\dot{b}\over a\,
b}-{{\dot{b}}^2\over b^2}-{\ddot{b}\over b}\right) =4 \Lambda_0\,.\nonumber\\[.2cm]
\label{enconext}
\end{eqnarray}
Equation (\ref{bir}) has the solution
\begin{equation} \dot{b}=a\, b\, k(t)\,,
\end{equation}
where $k(t)$ is an arbitrary function of time, then equation (\ref{iki}) becomes
\begin{equation} \left({a^{\prime\prime}\over a}+{b^{\prime\prime}\over
b}\right)-\left({1\over \rho}+2\,{b^\prime\over b}\right) \left({a^{\prime}\over
a}+{b^{\prime}\over b}\right)=\,{2\,q^2\over b^2\rho^4} \label{uc}\,.
\end{equation}
A solution of the full Einstein field equations can be found by following the method
given in the reference \cite{mashhoon} as follows:
\begin{eqnarray}
a={1-\displaystyle{\nu\,\lambda^2\over r^2}\over 1+\displaystyle{\lambda\over
r}+{\nu\,\lambda^2\over r^2}}\,\,,\qquad\qquad b={1\over W^{1/2}}\,{\lambda_0\over
\lambda }\,r\,(1+\displaystyle{\lambda\over r}+\displaystyle{\nu\,\lambda^2\over
r^2})\,\label{bbaa}
\end{eqnarray}
\vskip.5cm \noindent where $\displaystyle\nu={1\over 4}\left(1-{\eta_0\,^2\over
\lambda_0\,^2}\right), \quad\displaystyle\lambda(t)={\lambda_0\over f(t)}\,,\quad
W=(\alpha-\gamma r^2)(\delta r^2-\beta)$ \vskip.5cm \noindent and
$(\alpha\delta-\beta\gamma)>0$. $\alpha, \beta, \delta, \gamma, \eta_0$ are all real and
$\lambda_0>0$, $\nu\geq 0 $ are positive constants. Moreover, $f(t)$ is positive
arbitrary function of time and cosmological constant $\Lambda_0$\,.

\noindent Since the radial coordinate transformation does not change the co-moving
character of the metric, for the sake of the brevity, we used the transformation
\begin{equation}
r=\left({\alpha \rho^2+\beta\over \gamma \rho^2+\delta}\right)^{1/2}. \label{transform}
\end{equation}
Then, the physical quantities mass-energy density and pressure satisfying relations
(\ref{alti}), (\ref{bes}), can be written in following form
\begin{eqnarray}
&&\mu=-\Lambda_0+3\,\left({\dot{f}\over f}\right)^2+{192
f^2\left(\alpha\beta(-{\eta_0}^2+{\lambda_0}^2)+2 r(\alpha\beta+\delta\gamma
r^4)\lambda_0 f+4\delta\gamma r^6 f^2\right) \over N_0^4}\,\,,\nonumber\\[.2cm]\label{density}
&&p=\Lambda_0-{r^2(5(\eta_0^2-\lambda_0^2)-8 r\lambda_0 f+4 r^2f^2)\,\dot{f}^2\over f^2
N_1}-{2\,N_0\,\ddot{f}\over f N_1 }\nonumber\\[.2cm]
&&\qquad\quad+ {64 f^2 (\alpha r^2-\gamma^2)^3 (\beta r^2-\delta)^2N_2\over N_0^5 N_1
(\alpha\delta-\beta\gamma)^5 r^{10}} \label{pressure}
\end{eqnarray}
with
\begin{eqnarray}
&&N_0=-{\eta_0}^2+{\lambda_0}^2+4 r\lambda_0 f+4 r^2f^2\,,\quad\quad
N_1=\eta_0^2-\lambda_0^2+4 r^2 f^2\,\,,\nonumber\\[1.cm]
       &&N_2=\frac{1}{r^{14}\,{( \beta \,\gamma  - \alpha \,\delta  )
       }^4}\left(4\,f^2{{{\eta }_0}}^4{( \alpha  - r^2\gamma  ) }^6{( \beta  - r^2\,\delta  ) }^2
  ( 4f^2r^2 + {{{\eta }_0}}^2 - {{{\lambda }_0}}^2
  )\right.\nonumber\\[.2cm]
&& +{( \beta \,\gamma  - \alpha \,\delta  ) }^4 r^4\,\left( -\alpha \beta\eta_0^6 -
    {( 2\,f\,r + {{\lambda }_0} ) }^4\,
     ( 4\,f^2\,r^6\,\gamma \,\delta  - \alpha \,\beta \,{{{\lambda }_0}}^2
     )\right.
    \nonumber\\[.2cm]
&&+
    {{{\eta }_0}}^4\,( -4\,f^2\,r^2\,
       ( -3\,\alpha \,\beta  + r^2\,\beta \,\gamma  + r^2\,\alpha \,\delta )  +
       8\,f\,r\,\alpha \,\beta \,{{\lambda }_0} + 3\,\alpha \,\beta \,{{{\lambda }_0}}^2 )  \nonumber\\[.2cm]
       &&+
    {{{\eta }_0}}^2( 2fr + {{\lambda }_0} )
     \left( 8f^3r^5( - \beta \gamma    - \alpha \delta  +
          3\,r^2\,\gamma \,\delta  )\right. \nonumber\\[.2cm]
&&    \left. \left. \left.     +
       4\,f^2\,r^2( -4\alpha\beta  + r^2\beta\gamma  + r^2\alpha\delta  +
          r^4\gamma\delta ) \lambda_0 -
       10\,f\,r\,\alpha \,\beta \,{{{\lambda }_0}}^2 - 3\,\alpha \,\beta \,{{{\lambda }_0}}^3 \right)
    \right)\right)\nonumber\\[.2cm]
\end{eqnarray} and the total
charge of the fluid is given by
\begin{equation}
q={\eta_0 \over (\alpha\delta-\beta\gamma)^2 r^{3}}\, (\alpha-\gamma r^2)^{3/2}\,(\delta
r^2-\beta)^{3/2}\,.\label{charge}
\end{equation}
%\newpage
Consider the spacetime given by the line element (\ref{bbaa}) describes whole spacetime
then, by setting of arbitrary constants $(\alpha,\, \beta,\, \delta,\, \gamma,\,
\eta_0,\, \lambda_0,\, \nu )$ and function of time $f$ in special forms, well known
solutions of Einstein field equations such as RN-de Sitter and charged McVittie
solutions can be obtained.\\

\vskip.2cm \noindent {\bf  1) RN-de sitter solution.} \\The line element of the isotropic
RN-de Sitter spacetime in \cite{gao} is given  by
\begin{eqnarray}
ds^2=-{\displaystyle{\left[1-{m^2\over w^2 r^2}+{\tilde{q}^2\over w^2 r^2}\right]^2}\over
\displaystyle{\left[(1+{m\over w r})^2-{\tilde{q}^2\over w^2
r^2}\right]^2}}\,dt^2+w^2\left[(1+{m\over w r})^2-{\tilde{q}^2\over w^2
r^2}\right]^2(dr^2+r^2d\theta^2+r^2\sin^2\theta^2 d\phi^2)\label{dsrn}\nonumber
\end{eqnarray}
\begin{eqnarray}
{}
\end{eqnarray}
where $m$ is the mass and $\tilde{q}$ is the electric charge of the black hole and
$w=e^H$ is function of the Hubble parameter which is in general a function of time. If we
take
\begin{eqnarray}
f=w=e^{\sqrt{\Lambda_0/3}\,\,\,\,t},\, \lambda_0=2{m},\,\,\,\, \eta_0=2
\tilde{q},\quad\alpha=\delta=1,\,\quad\beta=\gamma=0\,\,\,\,\,
\end{eqnarray}
the line element (\ref{bbaa}) reduces to isotropic RN-de Sitter line element
(\ref{dsrn}).

In this configuration $\eta_0/2$ and $\lambda_0/2$ are considered as total charge and
mass of the black hole, respectively. Since $\dot{f}/ f=\sqrt{\Lambda_0/3}>0$\,,
therefore \\$\dot{b}/b>0$\,, then the spacetime expands in time  and
$H=const.=\sqrt{\Lambda_0/3}$ corresponds to the cosmological constant of RN-de Sitter
spacetime.  Furthermore, since $m\,$ (or $\lambda_0$) and $\tilde{q}$\,(or $\eta_0$) are
constants, $\mu=p=0$, are all zero as expected.

\noindent The spacetime (\ref{bbaa}) contracts only if $\dot{f}/ f $ is negative. If we
take $f=e^{-\sqrt{\Lambda_0/3}\,\,t}$, it will correspond to isotropic RN black hole in
anti-de Sitter spacetime. Here the physical quantities mass-energy density and the
pressure are also zero as the previous case.

The cosmological constant is the reason of the time evolution of the problem. Therefore,
if we take $\Lambda_0=0$, the contraction (expansion) of the spacetime disappears, it
becomes static.

\vskip.4cm \noindent {\bf 2) Charged McVittie-de Sitter solution.}\\
A perfect fluid solution of  Einstein's equations corresponding to the Schwarzschild
field embedded in a Robertson–Walker background is given by McVittie \cite{mcvittie}. In
expanding universe the McVittie solution represents a white hole, whereas  it represents
a black hole in contracting universe \cite{nolan}. In addition to the repulsive effect of
the expansion of the spacetime, the repulsive character of the electric charges (Coulomb
force) of the fluid  will be greater in small regions and will support formation of white
hole in the beginning in the McVittie-de Sitter spacetime.

If we consider ($\eta_0\neq 0\,$ $\beta=0$, $\alpha$ and $\delta$ positive) in
(\ref{bbaa}), the charged Mc Vittie-de Sitter solution is obtained and the isotropic
metric components become
\begin{eqnarray}
a&=&\frac{4\,\alpha\,r^2\,{f}^2 +
    ( \delta + \gamma\,r^2)  \,
     ( {{{\eta }_0}}^2 - {{{\lambda }_0}}^2
     ) }{4\,\alpha\,r^2\,{f}^2 +
    4\,\,r\,
     {\sqrt{\alpha(\delta + \gamma\,r^2) }}\,f\,
     {{\lambda }_0} -
    ( \delta + \gamma\,r^2)  \,
     ( {{{\eta }_0}}^2 - {{{\lambda }_0}}^2
      ) },\nonumber\\[.2cm]
       b&=&\frac{4\,\alpha\,r^2\,{f}^2 +
    4\,\,r\,
     {\sqrt{\alpha(\delta + \gamma\,r^2)}}\,f\,
     {{\lambda }_0} -
    ( \delta + \gamma\,r^2 ) \,
     ( {{{\eta }_0}}^2 - {{{\lambda }_0}}^2
       ) }{4\,\alpha\,r^2\,
   ( \delta + \gamma\,r^2 ) \,f}\,.\label{mcvittie1}
   \end{eqnarray}
The matter density and the pressure can be written as
   \begin{eqnarray}
    &&\mu=-\Lambda_0+3\,\left({\dot{f}\over f}\right)^2\nonumber\\[.2cm]
    &&-
\frac{128f^3\left( 2f
       {\left( \alpha  - r^2\gamma  \right) }^6
       {{{\eta }_0}}^4 +
      3r^5{\alpha }^4\gamma {\delta }^3
       {\left( 2fr + {{\lambda }_0} \right) }^3 +
      r^4{\alpha }^4{\delta }^3{{{\eta }_0}}^2
       \left( 2f\left( \alpha  -
            4r^2\gamma  \right)  -
         3r\gamma {{\lambda }_0} \right)  \right)
      }{{\alpha }^4{\delta }^2
    {\left( {{{\eta }_0}}^2 -
        {\left( 2fr + {{\lambda }_0} \right) }^2
        \right) }^4}
\,\nonumber\\[.2cm]
&&p=\Lambda_0 -{256\,f^4\,\delta\,\gamma\,r^6
       \over
    N_1\,N_0^2}+\left({\dot{f}\over
f^2}\right){5(\lambda_0^2-\eta_0^2)+8r\lambda_0\,f-4\,r^2\,f^2\over
    N_1}-2\left({\ddot{f}\over f}\right) {N_0^3\over
      N_1}\,\nonumber\\[.2cm]
\end{eqnarray}

and the charge becomes
\begin{equation}
q={\displaystyle{\eta_0\over \alpha^2\delta^{1/2}}}(\alpha-\gamma\,r^2)^{3/2}.
\end{equation}
By taking cosmological constant $\Lambda_0=0$, and electric charge $\eta_0=0$\,,
uncharged ordinary McVittie solution can be recovered
\begin{eqnarray}
&&a={-2\sqrt{\alpha}\,r f+\sqrt{\delta+\gamma r^2}\,\lambda_0\over 2\sqrt{\alpha}\,r
f+\sqrt{\delta+\gamma r^2}\,\lambda_0}, \qquad b={2\sqrt{\alpha}\,r f+\sqrt{\delta+\gamma
r^2}\,\lambda_0\over 4\alpha\,r^2 (\delta+\gamma r^2)\,f}\,.
\end{eqnarray}

\vspace{.4cm} \noindent Mass-energy density and the pressure of the fluid become
\begin{eqnarray}
&&\mu=3\left({\dot{f}\over f}\right)^2+{384 \lambda_0\,\delta\,\gamma\,r^5\,f^3\over(2 r
f-\lambda_0)^3(2 r f+\lambda_0)^3}\,,\nonumber\\[.2cm]
&&p={256 \,\delta\,\gamma\,r^6\,f^4\over(2 r f-\lambda_0)(2 r f+\lambda_0)^5} +
\left({\dot{f}\over f^2}\right) {5\lambda_0+8\lambda_0\,r\,f-4r^2\,f^2\over (2 r
f-\lambda_0)(2 r f+\lambda_0)}- 2\,\left({\ddot{f}\over f}\right) {(2 r
f+\lambda_0)^5\over (2 r f-\lambda_0)}\,.
\nonumber\\[.2cm]
\end{eqnarray}
Now let us consider the charged fluid is confined in a region in which dynamical
evolution of the system is described by the time dependent function $f(t)$ and  the
exterior spacetime to the confined matter is RN-de Sitter. Exterior and interior regions
separate the spacetime into two distinct parts such that they meet on the boundary
surface. To complete the solution, it is necessary to show that the distinct solutions
must satisfy boundary conditions on the boundary surface. In the following section we
will investigate matching conditions.

\section{Matching conditions}
Let us consider a spherical boundary surface which divides spacetime into two distinct
four-dimensional manifolds which admit $\Sigma$ as their boundaries at
$r_\Sigma=b\rho_\Sigma=\hat{r}_\Sigma=const.$.

Let $\hat t$, $\hat r$, $\hat\theta$, $\hat\phi$ be the Reissner-Nordstr\" om de Sitter
coordinates for the matter-free region. Then, the metric is
\begin{equation}
ds_+^2=-A\, d\hat t^2+A^{-1} d\hat r^2+\hat r^2(d\hat \theta^2+\sin^2\hat \theta\, d\hat
\phi^2)\label{newmet}
\end{equation}
where
$$A=1-{2 M\over \hat r}+{Q^2\over \hat r^2}-{\Lambda \over 3}\,\hat r^2$$
with the interior metric is in original form and given by
\begin{equation}
ds_-^2=-a^2\, dt^2+b^{2}\, d \rho^2+b^{2}\, \rho^2(d \theta^2+\sin^2 \theta\,
d\phi^2)\end{equation} where $a$ and $b$ are functions of coordinates $\rho$ and $t$.
Under the coordinate transformations
\begin{eqnarray}
\hat \theta= \theta,\qquad\qquad \hat \phi= \phi
\end{eqnarray}
motion of the boundary surface  can be given by the following equations
\begin{equation}
f^+:\hat{r}-\hat{r}_\Sigma(\hat{t})=0,\qquad\qquad f^-:\rho-\rho_\Sigma=0.
\end{equation}
To match the exterior spacetime with the interior spacetime  we use the Israel junction
conditions \cite{israel}. These conditions require the interior and exterior solutions of
the gravitational field equations to be joined smoothly up to a coordinate transformation
but the partial derivatives may change discontinuously across the boundary surface of the
matter.

\noindent Let $ds_\Sigma ^2$ be the line element of the boundary surface $\Sigma$,
$ds_+^2$ represent the exterior and $ds_-^2$ represent the interior spacetime line
elements. The junction conditions which state the equality of the first fundamental forms
and the discontinuity of the second fundamental forms can be given as
\begin{eqnarray}
ds^2_\Sigma=ds_-^2\vert_{\Sigma}=ds_+^2\vert_{\Sigma}\,,\nonumber\\[.2cm]
K^+_{ij}-K^-_{ij}-g_{ij}\,K = \tau_{ij}\label{form}
\end{eqnarray}
where $K=\,g^{ij}\left(K^+_{ij}-K^-_{ij}\right)$ and  $\tau_{ij}$ is the surface energy
momentum tensor. In case of vanishing surface energy momentum tensor i.e. for
$\tau_{ij}=0$, the discontinuity condition reduces to the equality of forms i.e.,
equality of the extrinsic curvatures
\begin{equation}
K_{ij}^ + = K_{ij}^ -\,\,.
\end{equation}
The continuity of the metric components in (\ref{form}) at $\rho=\rho_\Sigma$ on the
boundary surface gives the following relations
\begin{eqnarray}
&& \hat{r}_\Sigma\,=\rho_\Sigma\,b(t,\rho_\Sigma)\,,\qquad d\tau=a(t,\rho_\Sigma)\,dt\,,\\[.2cm]
&&d\tau=\sqrt{A(\hat{r}_\Sigma)-{1\over A(\hat{r}_\Sigma)}\left({d\hat{r}_\Sigma\over
d\hat{t}}\right)^2}\,d\hat{t},
\end{eqnarray}
and
\begin{eqnarray}
&&{d\,\hat{r}\over d\tau}=\rho\,{db\over d\tau}\,={\rho\over
a}\,{db\over dt}\,\vert_\Sigma, \\[.2cm]
&&\left({d\tau\over d\hat{t}}\right)^2={A^2\over A+(d\hat{r}/d
\tau\,)^2}\,\,\vert_\Sigma\,\,.
\end{eqnarray}
The non-zero second fundamental forms for the interior and the exterior regions are given
by
\begin{eqnarray}
&&K^-_{\theta\theta}=\sin^2\theta\,K^-_{\phi\phi}=\rho\,(b\,\rho)^\prime\,\label{secfunb}\\[.2cm]
%&&K^+_{\theta\theta}=\sin^2\theta\,K^+_{\phi\phi}=\displaystyle{\hat{r}\,A\,
%\over\displaystyle{\sqrt{A-{1\over A}\left({d\hat{r}\over d\hat{t}}\right)^2}}}\,\\[.2cm]
&&K^+_{\theta\theta}=\sin^2\theta\,K^+_{\phi\phi}=\frac{\hat{r}\,A}{\sqrt{A-\displaystyle{1\over
A}\left({\displaystyle{d\hat{r}\over d\hat{t}}}\right)^2}}\,\label{secfuni}
\end{eqnarray}
and
\begin{eqnarray}
&&K^{-}_{\tau\tau}=-{a^\prime\over a\, b}\label{in1}\\[.2cm]
&&K^{+}_{\tau\tau}={d\hat{r}\over d\tau}\,{d^2\hat{t}\over d\tau^2}-\,{d\hat{t}\over
d\tau}{d^2\hat{r}\over d\tau^2}+{3\over 2A}\,{\partial A\over\partial
\hat{r}}\,\left({d\hat{r}\over d\tau}\right)^2\,{d\hat{t}\over d\tau}-{A\over
2}\,{\partial A\over\partial \hat{r}}\,\left({d\hat{t}\over
d\tau}\right)^3\,\label{out1}\,.
\end{eqnarray}
By using the equality of the angular components of the second fundamental forms
(\ref{secfunb}, \ref{secfuni}) we obtain the following relations
\begin{eqnarray}
&&A\,\hat{r}\,{d\hat{t}\over d\tau}\,=\rho\,(b\,\rho)^\prime\,,\label{deriv}\\[.2cm]
&&A=\left({\rho\over b}\,\,{\partial b \over \partial \rho}+1\right)^2-{\rho^2\over
a^2}\,\left({\partial b\over \partial t}\right)^2\,.\label{an}
\end{eqnarray}
Since $b$ is  function of time dependent function $f$,  (\ref{an}) gives a condition that
$f$ should satisfy at any time $t$\,. Furthermore, the equality of the timelike
components of the extrinsic curvatures, (\ref{in1}) with (\ref{out1}), gives
\begin{eqnarray}
\displaystyle{\displaystyle(A\, d\hat{t}/d\tau)_{,\tau}\over d\hat{r}/d{\tau}}={1\over a
b}\,{\partial a\over \partial \rho}\,,
\end{eqnarray}
in other words it corresponds to $G_{t\rho}=0$ that is, no new information is obtained
from timelike components of $K_{\mu\nu}^\pm$.
%which gives $G_{t\rho}=0$\,.
%It is seen that time component of the extrinsic curvatures $K^\pm_{\,\,\tau\tau}$ do not
%contribute to get new relation.
Since the radial pressure of the fluid is zero on the boundary surface,   this condition
reduces to the continuity of the energy momentum tensors in radial direction
$T^{\rho}_{\rho}\,=T^{\hat{r}}_{\hat{r}}\,\vert_\Sigma$ which gives
\begin{equation}
p(t,\rho)-\Lambda_0={q^2(\rho)-Q^2\over  b^4\, \rho^4}-\Lambda\,.\label{press}\,
\end{equation}
For the sake of the generality we started with different cosmological constants for
interior and exterior regions, but the continuity of the energy momentum tensor in the
radial direction compels their equality ``$\Lambda_0=\Lambda"\,.$ Then, the electric
charge distribution  can be written as
$$q=\cases{\eta_0 W_\Sigma^{3/2}/\Delta^{2}
\rho_\Sigma^{3}=Q\qquad &$\rho\geq \rho_\Sigma$\cr \noalign{\vskip5pt} \eta_0 W^{3/2}/
\Delta^{2} \rho^{3}\qquad\qquad &$\rho < \rho_\Sigma$\cr}$$  where $Q$ is the total
electric charge of the fluid confined in the region $\rho\leq \rho_\Sigma$, and
$W_\Sigma$ is the value of $W$ at $\rho=\rho_\Sigma$ given by (\ref{bbaa}).

\section{Gravitational collapse}
%A solution of Einstein equations corresponding an isotropic spherical charged fluid
%(interior) admitting RN-de Sitter exterior is given in the preceding sections. In this
%part, we examine the gravitational collapse of the charged fluid and show what kind of
%singularities are allowed during collapse by the system. Furthermore, when the system
%allows their formation, the nature of the singularities and their relations between the
%electric charge and the cosmological constant are also investigated.

If the collapse phenomena allows their formation two types of singularities may form
during collapse: physical and spacetime singularities \cite{joshi}. Physical
singularities make physical quantities (such as mass-energy density, pressure) singular
and the space-time singularities make the metric components and the curvature indefinite.
In the gravitational collapse manner, among the spacetime singularities the shell
focusing and shell crossing singularities are being considered. The shell crossing
singularity occurs at distances where change of the radius of the fluid sphere in radial
direction is zero $R^\prime=0$ (with $R>0$), and the shell focusing singularity forms at
distances which make radius of the fluid sphere zero ($R\rightarrow 0$). The shell
crossing singularities can be considered weak with respect to the shell focusing
singularity in the gravitational collapse treatment \cite{singh}. Therefore, we are only
interested in the formation of the shell focusing singularity, i.e., $R\rightarrow 0$ as
$\rho\rightarrow 0\,$  with $R^\prime>0$\,.

In the literature many factors effective on the formation of the naked singularities are
examined {\cite{joshi},  \cite{joshi2}, \cite{goncalves}} and it is pointed out that, one
way of a singularity to be naked is to disturb the apparent horizon surface and delay its
formation \cite{joshi2}. According to the idea, if the trapped surface forms before the
singularity surface then, the singularity becomes hidden inside a black hole. Otherwise,
the trapped surface forms after the singularity surface and  the singularity becomes
naked. In another words, if the time period for the formation of the event horizon is
longer than the time period for the formation of the singularities, singularities become
bare and they can be seen by distant observer. This criterion is probably easy and
efficient, but it is not clear if this is always equivalent to naked singularity
formation and in a way it is coordinate dependent statement. In one coordinate system
these two timings may be related in a certain way, but may not be related in another
coordinate system. Therefore, the coordinate independent and a full proof condition "the
families of null geodesics come out of the singularity" should be examined \cite{joshi}.
We will give this analysis in the next section.

Let us write metric components of (\ref{bbaa}) in the radial coordinates $\rho$
explicitly to examine under which circumstances the situation corresponds to collapse,
expansion or bounce
\begin{eqnarray}
&&a={4
f^2(\alpha\rho^2+\beta)-\left(\lambda_0^2-\eta_0^2\right)\,(\gamma\rho^2+\delta)\over 4
f^2(\alpha\rho^2+\beta)+4f\,\lambda_0\sqrt{(\alpha\rho^2+\beta)(\gamma\rho^2+\delta)}+\left(\lambda_0^2-\eta_0^2
\right)\,(\gamma\rho^2+\delta)}\,,\nonumber\\[.2cm]
&&b=\displaystyle{4
f^2(\alpha\rho^2+\beta)+4f\,\lambda_0\sqrt{(\alpha\rho^2+\beta)(\gamma\rho^2+\delta)}+\left(\lambda_0^2-\eta_0^2
\right)\,(\gamma\rho^2+\delta)\over
4f\,(\alpha\rho^2+\beta)(\gamma\rho^2+\delta)}\,\nonumber\\[.2cm]\label{ab}
\end{eqnarray}
and define the physical radius $R$, the radius of 2-sphere
\begin{equation}
R=b\,\rho= \rho\,\displaystyle{4
f^2(\alpha\rho^2+\beta)+4f\,\lambda_0\sqrt{(\alpha\rho^2+\beta)(\gamma\rho^2+\delta)}+\left(\lambda_0^2-\eta_0^2
\right)\,(\gamma\rho^2+\delta)\over 4f\,(\alpha\rho^2+\beta)(\gamma\rho^2+\delta)}
\,.\label{phisrad22}
\end{equation}
As the radial coordinate $\rho\rightarrow 0$ the physical radius shrinks to zero. We know
that if the spacetime is in the isotropic form  the shear tensor is zero then it reduces
to
\begin{equation}
{\dot{R}\over R}={\dot{b}\over b}\,.\,\label{g2}
\end{equation}
By means of shear free property of the spacetime, the change of rate of the physical
radius that is, the expansion rate  of the spacetime is given by
\begin{equation}
\theta={3\over a}\,{\dot{R}\over R}=\,{3\dot{f}\over f}\,\,.\label{g1}
\end{equation}
This equality states the dynamics of this isotropic collapse problem is related to ratio
$\dot{f}/f$, and collapse (expansion) situation can only occur for the negative
(positive) values of $\dot{f}/f$.  Since $f$ is a positive function of time, only if
$\dot{f}<0\,$ $(>0)$ the negative (positive) expansion rate corresponds to contracting
(expanding) physical radius $R$ \,. If $\dot{f}<0$ changes its sign after a time period
then, it is called bounce. $\dot{f}=0$ will be static solution. The time evolution of the
physical radius is explicitly written by
\begin{equation}
\,{\dot{R}\over R}={a\,\dot{f}\over f}\,={\dot{f}\over f}\, {4
f^2(\alpha\rho^2+\beta)-(\lambda_0^2-\eta_0^2 )\,(\gamma\rho^2+\delta)\over
4f^2\,(\alpha\rho^2+\beta)+4f\,\lambda_0
\sqrt{(\alpha\rho^2+\beta)(\gamma\rho^2+\delta)}+(\lambda_0^2-\eta_0^2
)\,(\gamma\rho^2+\delta)}\,. \label{timet}
\end{equation}
from equations (\ref{g2}) and (\ref{g1}). By analyzing the time dependency of the $R$ we
see that four different situation can be obtained: collapse, expansion, stable and
bouncing cases.

In the literature it is shown that positive cosmological constant delays the formation of
the singularities \cite{madhav, shapiro}. By considering slowing down effect of the
positive cosmological constant on the collapse process let us take time dependent
function as $f=e^{-(c-\sqrt{\Lambda_0/3})\,t}$\, where $c$ is a positive constant. Since
$\dot{R}/R=-(c-\sqrt{\Lambda_0/3})\,a$, with $\sqrt{g_{tt}}=a> 0$, the expansion
(contraction) becomes dependent directly to the values of the cosmological constant
$\Lambda_0$. If $3c^2>\Lambda_0$\,the collapse will continue to a certain radius and be
stopped by the Coulomb's repulsive force by the charge and it will reach to the central
point for uncharged matter. If $3c^2<\Lambda_0$ radius will expand until it reaches to
the boundary of the exterior region that is to the apparent horizon stated by the
matching condition (\ref{an})\,. Furtheremore $3c^2=\Lambda_0$  corresponds to static
case.  In this example all three possible situations are obtained: expansion, crunch and
stable cases.

If the cosmological constant effects are dominant in the dynamics, it will be more
convenient to choose sample function as $f=e^{c\,t-\sqrt{\Lambda_0/3}\, t^2}$ ($t\geq
0$\,) to emphasize the cosmological constant dependency. In this model the change of the
radius with time $\dot{R}/R=(-c + 2 \Lambda_0\, t)\, a$, $a>0$, is positive in the
beginning and negative for late time $t$. It means that the radius of the fluid will
decrease with time and after a period of time, here for $t\geq c \sqrt{3/4 \Lambda_0}$\,,
the radius will start to increase so, "bounce" situation is obtained. $a=0$ case will be
examined in the following part in details.

\vskip.4cm \noindent {\bf  Physical singularities.}\\
When pressures are non-zero, dynamical evolutions, as allowed by the Einstein equations,
are equally important as the initial data is to determine the final fate of collapse.
\cite{joshi4}\,. Dust solution of the gravitational problem is highly important,  but the
isotropic form of the spacetime and set up of the problem does not give the dust solution
consistent with the conditions.

When their singularities are examined, it is seen that   matter and charge densities
(\ref{density}, \ref{charge}) are regular everywhere but the pressure which is subjected
to the the weak energy condition (\ref{enmomcon}) and satisfying the energy conservation
condition (\ref{encon})
\begin{equation}
p=-(\,\mu+{1\over 3}\,{\dot{\mu}\over a}\,\,{\dot{f}\over f} \, )\label{p1}
\end{equation}
diverges at the distance $\rho=\rho_s$ which makes the metric
component \\$a=\sqrt{g_{tt}}=0$
\begin{eqnarray}
4f^2(\alpha\rho^2+\beta)-\displaystyle{(\lambda_0^2-\eta_0^2)}\,(\gamma\rho^2+\delta)=0,\,\,\,{\rm
or}\quad \rho_s=\sqrt{-4\beta\,f^2+\delta(\lambda_0^2-\eta_0^2)\over \,\,4\alpha\,
f^2-\gamma(\lambda_0^2-\eta_0^2)} \,.\label{singdist}
\end{eqnarray}

At $\rho_s$ the physical region is split into two parts i.e. matter part is confined in
$\rho\leq \rho_s$ and exterior part starts from $\rho> \rho_s$ where $a$ is non-negative.

One may think that the physical singularity starts from the origin which makes $\rho_s=0$
(\ref{singdist}).  But it can not be allowed due to Coulomb's repulsive force of the
fluid, or the fluid can only be compressed to a radius at which Coulomb interactions
balance the the gravitational collapse effects. This restriction can be seen from
conservation relation (\ref{encon}), that is time derivative of $\mu$ must be positive.
By taking the charge parameter limit highly big values ($\eta_0\rightarrow \infty$ ) in
$\dot{\mu}$,  it gives us a relation to be satisfied by $f$
\begin{equation} {\dot{f}\over f} \left({\dot{f}\over f}\right)^.\geq 0\,.
\end{equation}
Since $f>0$ and $\dot{f}/f<0$ for all $t$, that is $f$ is decreasing function with time,
then the equation is always negative or zero. Zero case corresponds to static solution
but for the collapse situation we get only negative values. Thus, existence of the
electric charge violates the conservation of the energy momentum for $\rho\rightarrow
0$\,.  In other words, The conservation of the energy-momentum does not allow formation
of the shell focusing singularity. For uncharged fluid $\eta_0=0$, the central
singularity can be reached and $f=\sqrt{\delta\lambda_0^2/4\beta}$ gives us shell
focusing singularity formation period $f$.

For example, if we take time dependent function for neutral matter as
$f=e^{-(c-\sqrt{\Lambda/3})\,t}$\,, $c>0$\, with $\beta=\gamma=0$ in (\ref{singdist})\,,
the pressure will be singular at the radial distance $\rho_s \sim$ $
f^{-1}=e^{(c-\sqrt{\Lambda_0/3})\,t}$ which means singularity will form later than
$\Lambda_0$ free case but the radius of the singularity surface will be greater than that
of $\Lambda_0$.

\vskip.5cm \noindent {\bf Spacetime singularities.}\\
The Kretschmann scalar which is the square of the Riemann tensor and defined by
$K=R_{\mu\nu\lambda\sigma}\,R^{\mu\nu\lambda\sigma}$ gives the essential, coordinate
independent singularities of the spacetime. For the spherically symmetric isotropic
spacetime it can be given as
\begin{eqnarray}
K&&={1\over a^6\,b^8\,\rho^2}\,\left[12
a^4\rho^2{b^\prime}^2(a^2{b^\prime}^2+b^2{a^\prime}^2)
+8b^2a^4(3a^2{b^\prime}^2+b^2{a^\prime}^2)\right.\nonumber\\[.2cm]
&&+12b^4\rho^2{\dot{b}^2}(a^2{\dot{b}^2}+b^2{\dot{a}^2})
+8a^2b^4\rho^2\dot{a}\dot{b}({a^\prime}{b^\prime}+ba^{\prime\prime})
-8a^2b^2\rho^2{\dot{b}^2}(a^2b^{\prime\,2}+2b^2a^{\prime\,2})\nonumber\\[.2cm]
&& -16a^2b^3\rho{a^\prime}{\dot{b}}(2a\rho{\dot{b}}b^\prime-b^2\dot{a})
+8a^4b^3\rho{a^\prime}{b^\prime}(2{a^\prime}-\rho b^{\prime\prime})+32a^4b^3\rho{b^\prime}{\dot{b}}(\rho{\dot{b}}^\prime-\dot{b})\nonumber\\[.2cm]
&&+12ab^6\rho^2\ddot{b}(a\ddot{b}-2\dot{a}\dot{b})+ 8a^6b^2\rho
b^{\prime\prime}(\rho+2b^\prime)+4a^3b^4\rho^2a^{\prime\prime}(aa^{\prime\prime}-2b\ddot{b})\nonumber\\[.2cm]
&&\left.-16a^4b\rho^2b^{\prime\prime}(a^2b^{\prime\,2}+b^2\dot{b}^2)-8a^3b^4\rho\ddot{b}a^\prime(\rho
b^\prime+2 b a^\prime)+16 a^3 b^4 b\rho^2
\dot{b}^{\prime}(2\dot{b}a^\prime-a\dot{b}^{\prime})\right]\,.\nonumber\\[.2cm]
\end{eqnarray}
$K$ has polynomial singularities in $a, b, \rho$ and divergent as
$\sqrt{g_{tt}}=a\rightarrow0  $ (physical singularity), or $\rho\rightarrow 0$\,,
$R=b\rho\rightarrow 0\,$ (central singularity)\,.

If the future directed non-spacelike (timelike or null) curves terminate in the past at
the singularity then the singularity is called naked otherwise it is covered. The
procedure is coordinate-free method. To clarify the nature of the singularities the
future directed non-spacelike geodesics are examined, specially null geodesics
\cite{joshi, singh}\,. Outgoing radial null geodesics of the isotropic spacetime
(\ref{flrwline}) are given by
\begin{equation}
{dt\over d\rho}={b\over a}\,\,.
\end{equation}
If null geodesic equation is written in terms of physical radius $R$ and $u=\rho^\alpha$
we get
\begin{equation}
{dR\over du}={R^\prime\over \alpha \rho^{\alpha-1}}\left(1+{b\over a}{ \dot{R} \over
R^\prime}\right)\,.
\end{equation}
The singularity is naked if the null geodesics terminate in the past at the singularity
with positive finite value and it is hidden or covered if the limit
\begin{equation}
\lim_{\rho\rightarrow \rho_s}{dR\over du}=\lim_{\rho\rightarrow \rho_s}{R^\prime\over
\alpha \rho^{\alpha-1}}\left(1+{b\over a}{\dot{R} \over R^\prime}\right)\,\label{limit}
\end{equation}
is negative. As stated before, the existence of the electric charge restricts the
formation of the central singularity, only uncharged matter collapse ($\eta_0=0$) ends
with central singularity. If all constants about the fluid defined in (\ref{bbaa}) are
non-zero except $\eta_0=0$ and for $a\neq 0$ singularity starts from origin. The second
term in the parenthesis becomes zero as $\rho_s\rightarrow 0$ since $R^\prime>0$ and
$\dot{R}=\rho\,\dot{b}$. Therefore, limit becomes positive and equal to 1, then the
central singularity is naked.  If we take $\eta_0=\beta=\gamma=0$ in (\ref{singdist}) at
which  $a=0$\,  singularity forms at $\rho_s=\sqrt{\delta\lambda_0^2/4\alpha f^2} $\,. In
this limit the (\ref{limit}), therefore the nature of the singularity  becomes parameter
dependent.

In this case the limit becomes related to the sign of the time derivative of $f$ and the
constant $\alpha$ as \,\,$\rm{sgn}(\dot{f})/\rm{sgn}(\alpha)$\,. For  ($\alpha>0,
\dot{f}>0$) or ($\alpha<0, \dot{f}<0$ with $\delta<0$), limit becomes positive and
therefore the central singularity becomes naked. But, for ($\alpha>0, \dot{f}<0$) or
($\alpha<0, \dot{f}>0$ with  $\delta<0$) limit is negative therefore, the central
singularity is covered.  It is possible to give a lot of example such $f$ functions so
that ${\rm sgn}\dot{f}$ changes with time during process other than
$f=e^{c\,t-\sqrt{\Lambda_0/3}\, t^2}$ where "bounce" situation occurs. Otherwise the
process will be collapse or expansion.

Furthermore, it should be also emphasize that the physical singularity ($a\rightarrow 0$
singularity) coincides with the apparent horizon for the extremal case $\lambda_0=\eta_0$
and $\beta=0$\,. Apparent horizon is the boundary surface of the trapped regions and
makes (\ref{an}) zero
\begin{equation}
A=b^{-2}{R^{\,\prime}}^{2}-a^{-2}{\dot{R}}^{2}=0\,.
\end{equation}

\section{Conclusion}

In this work spherical symmetric charged solution of Einstein field equations is given in
the presence of cosmological constant. The matter is considered as ideal fluid and
subjected to the weak energy condition. Two specific examples of this type of solution
isotropic RN-(anti)de Sitter and charged Mc Vittie-de Sitter solutions are given. When
the matter is confined in a region, the exterior spacetime exterior is taken RN-de Sitter
and to complete analysis the matching conditions are examined. In these calculations, in
the name of the generality, we started with the different cosmological constant for the
interior $\Lambda_0$ and exterior $\Lambda$ regions but the junction conditions,
continuity of the energy-momentum tensor in the radial direction gives their equality.

In the reference \cite{mashhoon}, spherical symmetric gravitational collapse  of charged
fluid is studied in black hole formation point of view, that is, strong cosmic censorship
hypothesis is considered. In this work, by considering weak cosmic censorship hypothesis
we see that besides formation of the black hole the process allows formation of the naked
singularity and the existence of the cosmological constant permits bouncing situations as
well. The singularity structure analysis is done by using coordinate free null geodesic
method. The initial data  about the matter (energy density, pressure and the electric
charge) therefore, constants about the solution determine the final fate of the collapse.
Existence of the electric charge prevents formation of the central singularity and the
cosmological constant causes  bouncing situation for both charged and neutral matters.
Uncharged matter distribution allows formation of the central singularity and it will be
naked.

The results are compatible with the results given in the literature. For further studies,
the gravitational collapse phenomena can be studied for the spacetimes other than
isotropic form.

\noindent{\bf{Acknowledgements.}} Author would like to thank Prof. B.Mashhoon and Prof.
P. Joshi for valuable suggestions and also would like to thank Prof. M. Horta\c csu for his
kind interest. This work is partially supported by TUBITAK.

\end{document}